\documentclass[aps,prl,superscriptaddress,twocolumn,floatfix,a4paper]{revtex4}

\usepackage{graphicx,graphics,epsfig}   
\usepackage{dcolumn}    
\usepackage{bm}         
\usepackage{amsmath}    
\usepackage{verbatim}   
\usepackage{color}      
\usepackage{subfigure}  
\usepackage{times,natbib}
\usepackage{amsmath,amsfonts,amssymb,graphics,graphicx,epsfig,color,times,natbib}

\newcommand{\bra}[1]{\langle#1| }
\newcommand{\ket}[1]{|#1\rangle }

\newcommand{\ba}{\begin{eqnarray}}
\newcommand{\ea}{\end{eqnarray}}

\begin{document}

\title{Testing the Structure of Multipartite Entanglement with Bell inequalities}

\author{Nicolas Brunner}
 \affiliation{H H Wills Physics Laboratory, University of Bristol, Bristol BS8 1TL, United Kingdom}
\author{James Sharam}
 \affiliation{Department of Mathematics, University of Bristol, Bristol BS8 1TW, United Kingdom}
\author{Tamas V\'{e}rtesi}
 \affiliation{Institute of Nuclear Research of the Hungarian Academy of Sciences, H-4001 Debrecen, P.O. Box 51, Hungary}

\begin{abstract}
We show that the rich structure of multipartite entanglement can be tested following a device-independent approach. Specifically we present Bell inequalities for distinguishing between different types of multipartite entanglement, without placing any assumptions on the measurement devices used in the protocol, in contrast with usual entanglement witnesses. We first address the case of three qubits and present Bell inequalities that can be violated by W states but not by GHZ states, and vice versa. Next, we devise 'sub-correlation Bell inequalities' for any number of parties, which can provably not be violated by a broad class of multipartite entangled states (generalizations of GHZ states), but for which violations can be obtained for W states. Our results give insight into the nonlocality of W states. The simplicity and robustness of our tests make them appealing for experiments.
\end{abstract}


\maketitle


Entanglement \cite{rmp_horo} plays a central role in quantum theory, in particular in the context of quantum information. Consequently, its characterization and detection are important issues, which have been extensively studied recently \cite{review_toth}.

While the structure and detection of entanglement is relatively well understood in the bipartite case \cite{rmp_horo,review_toth}, much less is known in the multipartite case (i.e. for three or more systems). Given the importance of multipartite entangled states in the context of quantum computation \cite{cluster}, quantum simulations \cite{lloyd}, quantum metrology \cite{pezze}, and many-body systems \cite{mps}, an intense research effort has been recently witnessed in this direction. Sophisticated techniques, such as entanglement witnesses, have been developed (see e.g. \cite{huber2} and references therein). This line of research is also important from an experimental perspective, given recent progress achieved with photons \cite{bourennane}, trapped ions \cite{blatt}, and superconducting qubits \cite{neeley}.

The central aspect of multipartite entanglement---which makes its study a highly nontrivial problem---is its rich and complex structure. Three (or more) quantum systems can be entangled in very different ways \cite{dur}. The most celebrated example is the case of three qubits, which feature two inequivalent classes of tripartite entangled states, namely Greenberger-Horne-Zeilinger (GHZ) \cite{ghz} and W states \cite{dur}. 

Importantly, different types of multipartite entanglement can have completely different features, for instance from the point of view of strength of correlations, or robustness to noise and losses. In general this greatly affects their information theoretic properties.
For instance, GHZ states, which are highly entangled but very fragile with respect to losses, are essential for many quantum information-theoretic tasks, such as quantum secret sharing \cite{hillery} and quantum teleportation \cite{zhao}. On the other hand, W states, which are much less entangled but highly robust against losses, are central to the physics of the interaction of light and matter \cite{dicke}, in particular for quantum memories \cite{q_memory}. More generally, it is also known that too much entanglement hinders computational power \cite{gross}.

It is therefore important to understand how different types of multipartite entanglement can be characterized and distinguished from each other.
Recently Schmid et al. \cite{schmid} and Huber et al. \cite{huber} discussed this problem, introducing entanglement witnesses that allow one to distinguish between important classes of qubit states. 

However, entanglement witnesses have an important drawback, in particular from the experimental point of view, which is that their derivation is based on  assumptions which are difficult (if not impossible) to meet in practice. In particular, the violation of an entanglement witness can be considered as a conclusive proof of entanglement only if the measurement operators are precisely characterized. This is indeed impossible in practice where (small) alignment errors are inevitable. Importantly, such errors can have dramatic consequences, as recently shown in Refs \cite{bancal}; uncertainties on the measurements make it possible (in principle) for a separable state to lead to a violation of the witness, thus rendering the test inconclusive. Therefore, it is desirable to devise methods for both witnessing multipartite entanglement and for testing its structure, which do not rely on any assumptions about the measuring devices used in the protocol. 

The first problem has been addressed recently via the introduction of device-independent witnesses for multipartite entanglement \cite{bancal}. Here, we address the second question, and show how the structure of multipartite entanglement can be tested in a device-independent manner. 
To do so, we present Bell inequalities which (i) are tailored for detecting the nonlocality of certain multipartite entangled states, and (ii) can provably not be violated by other classes of multipartite entangled states. Hence the violation of our inequalities allows one to discriminate between various types of multipartite entanglement. Importantly this is achieved without placing any assumptions on the measurement devices used in the protocol, since the violation of a Bell inequality is a statement based only on the observed statistics. Here, we start by addressing the case of three qubits, and show how one can distinguish between GHZ and W entanglement. 
Then we move to the more general case and introduce a class of Bell inequalities which can provably not be violated by a large class of multipartite entangled states (in particular generalized GHZ states). The key feature of our inequalities is that they are based on sub-correlations, i.e. each term of the Bell polynomial involves only a subset of the parties. We derive a family of 'sub-correlation Bell inequalities' for an arbitrary number of parties, and show that they can be violated by W states. Finally we describe a method for deriving useful sub-correlation Bell inequalities and apply it to simple cases. Generally, our results give insight into the nonlocality of W states \cite{cabello}, which is still poorly understood.


{\bf \emph{Preliminaries.}} We will focus on two inequivalent classes of multipartite entangled states of $N$ systems: (i) W states:
\ba\label{W}	\ket{W_N} = \frac{1}{\sqrt{N}} \left( \ket{0 \hdots 0 1} +   \ket{0 \hdots 1 0} + \hdots + \ket{1 0 \hdots  0} \right)	\ea
and (ii) generalized $d$-dimensional GHZ states:
\ba\label{schmidt}	\ket{\Psi_N^d} =  \sum_{j=0}^{d-1} \alpha_j \ket{j  }^{\otimes N} \ea
where $\sum_{j=0}^{d-1} |\alpha_j|^2=1$. We will also refer to GHZ states:
\ba\label{ghz}	\ket{\text{GHZ}_N} =  \frac{1}{\sqrt{2}} ( \ket{0  }^{\otimes N} +  \ket{1  }^{\otimes N}). \ea
Here, following \cite{huber}, entanglement classes are defined by equivalence under local unitaries and permutation of the parties \footnote{Note that this classification is different from that based on local operations and classical communication (LOCC), see e.g. Acin \emph{et al.} Phys. Rev. Lett. {\bf 87}, 040401 (2001). The latter is not adequate here, since CC is a nonlocal resource.}. Both classes are robust to imperfections, however in a very different manner. For instance, GHZ states are maximally entangled and thus robust against noise. On the other hand, W states are robust against losses; tracing out $k\ll N$ particles leaves the state barely unchanged. This is not the case for generalized GHZ states which are extremely fragile to losses; indeed it suffices to trace out a single subsystem to destroy all entanglement. This difference will play a crucial role later on.

The central tool here will be Bell-type inequalities. We shall focus on multipartite Bell polynomials which are symmetric under any permutation of the parties. Each party can perform two possible measurements featuring binary outputs. Therefore the following notation will be useful; we write
\ba\label{sym}	 \left[ \alpha_0 \mbox{ } \alpha_1 ; \mbox{ } \alpha_{00} \mbox{ } \alpha_{01} \mbox{ } \alpha_{11} \right]  &\equiv&  \alpha_0(A_0 + B_0) + \alpha_1 (A_1 + B_1)  \nonumber  \\ \nonumber
		& &+ \alpha_{00}A_0B_0 + \alpha_{01}(A_0B_1 + A_1B_0) \\ & &+ \alpha_{11}A_1B_1 \ea
where $A_i=\pm 1$ denotes the outcome of measurement setting $i=0,1$ of party $A$. Likewise for $B_i$. The extension to more parties is straightforward. For instance, for $N=3$ parties, the Mermin inequality \cite{mermin}, usually given by
\ba M_3 = A_0B_0C_0 - A_0B_1C_1 - A_1B_0C_1 - A_1B_1C_0 \leq 2 \ea
now reads
\ba\label{mermin} M_3 =  \left[ 0 \mbox{ }0  \mbox{ }; 0\mbox{ } 0\mbox{ } 0\mbox{ } ; 1\mbox{ } 0 \mbox{ }-1\mbox{ }0 \right] \leq 2 . \ea
Similarly Svetlichny's inequality \cite{svetlichny} will now be written as
\ba\label{svet} S_3 =  \left[ 0 \mbox{ }0  \mbox{ }; 0\mbox{ } 0\mbox{ } 0\mbox{ } ; 1\mbox{ } 1 \mbox{ }-1 \mbox{ }-1 \right] \leq 4 . \ea
while the generalized Mermin-Ardehali-Belinskii-Klyshko \cite{mermin,ABK} (MABK) Bell inequality is given by
\ba\label{mermin_N} M_N =
\begin{cases}
[0 ... 0; \underbrace{1 \,\, 0  \,\,-1 \,\, 0 \,\, 1 \,\, 0 \,-1 \hdots}_{N+1} ] \leq 2^{\frac{N-1}{2}}& \text{odd N} \\
[0 ... 0; \underbrace{1 \,\, 1  \,\,-1 \,\,-1 \,\, 1 \,\, 1 \,-1  \hdots}_{N+1} ] \leq 2^{\frac{N}{2}} & \text{even N.}
\end{cases} \ea


{\bf \emph{Case of 3 qubits.}} We show how to discriminate in a device-independent way between the states $\ket{W_3}$ and $\ket{\text{GHZ}_3}$.

First, we consider the Mermin inequality \eqref{mermin}. Here the state $\ket{\text{GHZ}_3}$ can lead to maximal violation (i.e. $M_3=4$), which corresponds to the GHZ paradox \cite{ghz}. However, only much lower violations are possible for the $\ket{W_3}$ state. Following the method of \cite{dimH}, based on semidefinite programming, we could prove that $M_3 \leq 3.079$ for $\ket{W_3}$; details can be found in Appendix A. Thus a violation of the inequality $M_3\leq 3.079$, which is possible using $\ket{\text{GHZ}_3}$, certifies that no $\ket{W_3}$ states were used. The robustness of this effect is best represented using the resistance to noise, which is here found to be $w_{\text{GHZ}_3}\sim 77\%$. Physically this means that unless a fraction of (at least) $(1-w_{\text{GHZ}_3})$ of white noise is mixed to $\ket{\text{GHZ}_3}$ the inequality $M_3\leq 3.079$ will be violated.

Next, we study the converse task. Let us consider the following inequality
\ba\label{B}	B = S_3 + R_2 = \left[ \mbox{ } 0 \mbox{ } 0 \mbox{ } ; \mbox{ } 0 \mbox{ } 1 \mbox{ } 0  \mbox{ }; \mbox{ } 1 \mbox{ } 1 \mbox{ } -1 \mbox{ } -1 \right] \leq 6. \ea
Thus $B$ is simply the Svetlichny polynomial $S_3$ \eqref{svet}, to which we have added some 2-party correlation terms, denoted $R_2$.

On the one hand, it turns out that inequality \eqref{B} can be violated by the state $\ket{W_3}$. The largest violation is $B\simeq 7.26$ (see Appendix B), leading to a resistance to noise of $w_{W_3}\simeq 82.7\%$. On the other hand, inequality \eqref{B} can provably not be violated by generalized GHZ states \eqref{schmidt} with $d=2$; we conjecture that inequality \eqref{B} holds for all states \eqref{schmidt}, which we checked numerically for $d\leq 10$. Below we sketch the proof for the state $\ket{\text{GHZ}_3}$; the full proof for 2-dimensional generalized GHZ states can be found in Appendix C. Finally, note that inequality \eqref{B} holds for any bipartition, that is, for any configuration in which two (out of three) parties can group and have any no-signaling correlations.

\emph{Sketch of the Proof.} Consider the decomposition of the polynomial $B = S_3 + R_2$. The intuition behind the proof is the following. On the one hand, the state $\ket{\text{GHZ}_3}$ can lead to a high violation of the Svetlichny inequality ($S_3 \leq 4\sqrt{2}$, see \cite{bancal2}) using qubit measurements in the XY plane (i.e. the equator) of the Bloch sphere. In that case however, the state will exhibit no 2-party correlations, leading to $R_2=0$. Thus overall we get no violation. On the other hand, the GHZ state has perfect 2-party correlations for measurements along the Z axis. Thus one can achieve $R_2=6$ with such measurements. However, in this case, there will be no 3-party correlations, thus leading to $S_3=0$. Overall, we get no violation again.

Based on this intuition we build the proof as follows. First we write the measurement operators as $A_j = \cos{\theta_j} XY + \sin{\theta_j} Z$
where $XY$ denotes a Pauli operator on the XY plane, i.e. of the form $\cos\phi X+\sin\phi Y$; here X, Y, Z are the usual Pauli matrices. One can then show that the $S_3$ part of the polynomial is upper bounded by an expression depending upon the cosine of the angles $\theta_j$, whereas the $R_2$ part of the polynomial is upper bounded by an expression of the sine of the angles $\theta_j$. After trigonometric massaging, it follows that $B\leq 6$. $\hfill \blacksquare$


{\bf \emph{General case; Sub-correlation Bell inequalities.}} We now move to the case of an arbitrary number of parties. Our main tool will be sub-correlations Bell inequalities (SCBI). These are based on polynomials which do not include any full $N$-party correlation terms. Here we focus on SCBI which are symmetric (under permutation of the parties).

The motivation for studying such inequalities comes from the fact that different types of multipartite entanglement are affected differently by losses of particles. On the one hand, losses barely affect W states. For $N$ large enough, we have

\ba\label{W_losses} \text{tr}_N(\rho_{W_N}) = \frac{1}{N} \ket{0}\bra{0}^{\otimes N-1} + (1-\frac{1}{N}) \rho_{W_{N-1}} \approx \rho_{W_{N-1}} \ea
where $\rho_{W_N} = \ket{W_N} \bra{W_N}$, which suggests that W states can violate SCBI. On the other hand, one would intuitively expect that states which are not entangled anymore after having traced out one particle, will never violate a sub-correlation Bell inequality. Although this may not be true in general \cite{lars}, we will see below that it holds nevertheless for a wide class of multipartite entangled states, in particular for all generalized GHZ states \eqref{schmidt}.
	\begin{table}[b!]
	\caption{Violations (given as resistance to white noise) for $\ket{W_N}$ states of the SCBI \eqref{B_N}.}\label{table:quantum-violations}
	\centering
	\begin{tabular}{ c | c c c c c c c c c c }
	\hline \hline
	$N$ &  4 & 5 & 6 & 7 & 8  & 10 & 12 & 15 & 20 & 40   \\
	\hline  \\
	$w_{W_N}$ & 1 &  0.891 & 0.831 & 0.792 & 0.765  & 0.730 & 0.709 & 0.688 & 0.669 & 0.642 \\
	\hline \hline
	\end{tabular}
	\end{table}

In order to show that a given state $\rho_N$ does not violate any SCBI, it is sufficient to ensure that all of its $N$ reduced states, obtained by tracing out one party, are compatible with a separable state $\sigma_N$, i.e. $\text{tr}_k(\rho_N)=\text{tr}_k(\sigma_N)$ for all $k\in\{1,...,N\}$. In this case, there can be no Bell violation, since the separable state $\sigma_N$ provides a local model. For any state of the form \eqref{schmidt}, a separable state satisfying the above condition is given by $\sigma_N = \sum_{j=0}^{d-1}|\alpha_j|^2 \ket{j \hdots j}\bra{j \hdots j}$. Thus generalized GHZ states can never violate any SCBI.

We now present a family of sub-correlation Bell inequalities for an arbitrary number of parties $N$, which can be violated by W states \eqref{W}. This construction is based on the MABK polynomials, and is given by

\ba\label{B_N}	B_N & = M_{N-1}^A + M_{N-1}^B + \hdots + M_{N-1}^N  \leq N 2^{\lceil \frac{N-2}{2}\rceil }\ea
where $M_{N-1}^k$ denotes the polynomial in \eqref{mermin_N} involving all parties except party $k$. Given that $M_{N-1}^k \leq  2^{\lceil \frac{N-2}{2}\rceil }$ (see \eqref{mermin_N}), it follows that $B_N$ is upper bounded by $  N 2^{\lceil \frac{N-2}{2}\rceil }$, the sum of all the local bounds.
It turns out that this bound is tight, i.e. there exists a local model (for all $N$ parties) which achieves the local bound of each MABK polynomial in \eqref{B_N}.

Next we study the achievable violations of inequality \eqref{B_N} for states $\ket{W_N}$. We start by writing the W density matrix in the basis of Pauli matrices $\{\openone,X,Y,Z\}$:

\ba \label{rho_W} \rho_{W_N} &=&   \frac{1}{N2^N}  \text{sym} \bigg[  \, \sum_{k=0}^N (N-2k)  \openone^{\otimes N-k} Z^{\otimes k} \\\nonumber
& & \quad \quad \quad \quad \,\,\,+ 2  ( \openone + Z)^{\otimes N-2} (XX+YY)   \bigg]  \ea
where $\text{sym}[x]$ means the symmetrization of expression $x$.

Noting the XY symmetry in the above expression, we focus here on measurements in the XZ plane of the Bloch sphere $A_j = c_j Z +s_jX$
with $c_j=\cos{\theta_j}$ and $s_j=\sin{\theta_j}$ ($j=0,1$). Also, we assume that all parties perform the same two measurements, i.e. $A_j=B_j=C_j$ and so on.

We can now derive an expression for a full correlation term $\mathcal{T}_k^N=A_x\otimes B_y \otimes \hdots \otimes N_z$ featuring $N-k$ settings '0' and $k$ settings '1':
\ba\label{fullC} \text{tr}(\rho_{W_N} \mathcal{T}_k^N) = -c_0^{N-k} c_1^{k} + \frac{2}{N}  \bigg[  \left( \begin{array}{c} k \\ 2 \end{array} \right) c_0^{N-k} s_1^2 c_1^{k-2}     \\\nonumber  + k(N-k) c_0^{N-k-1} s_0 s_1 c_1^{k-1}   \\\nonumber + \left( \begin{array}{c} N-k \\ 2 \end{array} \right) c_0^{N-k-2} s_0^2 c_1^{k}    \bigg]    \ea
Inequalities \eqref{B_N} involve only correlation terms of $(N-1)$ parties. Using the equality in \eqref{W_losses}, we can evaluate a term $\mathcal{T}_k^{N-1}$ featuring $N-1-k$ settings '0' and $k$ settings '1':

\ba \text{tr}(\rho_{W_N} \mathcal{T}_k^{N-1}) &=& \frac{1}{N} \text{tr}(\ket{0}\bra{0}^{\otimes N-1}\mathcal{B}_k^{N-1}) \\\nonumber
&&+(1-\frac{1}{N}) \text{tr}(\rho_{W_{N-1}} \mathcal{B}_k^{N-1})\ea
The first line of the above expression is then given by $c_0^{N-k}c_1^k/N $, while the second line can be evaluated using \eqref{fullC}.

Thus, we get an expression for the value of $B_N$ for states $\ket{W_N}$ which is a function of only two angles ($\theta_{0,1}$). Using numerical optimization, we investigated W violations up to $N=40$. The results are presented in Table~1, and are presented in terms of resistance to noise, which is given here by $w_{W_N} = N 2^{\lceil \frac{N-2}{2}\rceil } / \text{tr}(\rho_{W_N} B_N)$. The violation of \eqref{B_N} thus clearly increases with $N$; we conjecture that for all $N\geq 5$, the W state leads to a Bell violation. Also, for $N=5$, we could check numerically that our choice of measurement settings is optimal.

\emph{Finding sub-correlation Bell inequalities.} Further relevant SCBI can be found using the geometrical approach to nonlocal correlations. In the latter, Bell inequalities correspond to the facets of the polytope of local correlations. The extremal points of this polytope are given by local deterministic strategies---in which each party outputs deterministically for all measurements---and are thus easily characterized. Then, using appropriate software, it is possible to find the facets of a given polytope, and thus retrieve Bell inequalities.

In order to find SCBI, we follow the above approach, but forget about full ($N$-party) correlation terms, thus effectively projecting the initial local polytope in a subspace of smaller dimension. Here we have followed this method in the symmetric approach (i.e. focusing on inequalities which are symmetric under permutation of the parties), which greatly simplifies the problem \cite{jd_sym}, and considering two measurements per party, all with binary outcomes. This allowed us to find SCBI for $N=4,5$. Here we present two of them, which are relevant from the point of view of quantum violations:

\ba\label{I4} I_4 &=& [\,-1\,\, -1\,;\, -2 \,\,0\,\, -2 \,;\, -2\,\, 1 \,\,1\,\, -2\,] \leq 8   \\\nonumber
 I_5 &=& [\,0\,\, 0\,;\, -2 \,\,0\,\, -1 \,;\, 0\,\, 0 \,\,0\,\, 0\,; \, -4 \,\,0 \,\,2\,\, 0\,\, 1\,] \leq 15 \ea
Note that here we have used the notation \eqref{sym}, omitting the coefficients for full correlation terms since the latter are always zero for SCBI. For these inequalities, we have found large violations for the W state; for $I_4$ we get $w_{W_4}\simeq0.707$, while for $I_5$ we get $w_{W_5}\simeq0.536$ (see Appendix B).

Finally, it is worth mentioning that a phenomenon of frustration occurs for both of these inequalities. Specifically, each symmetric SCBI $J_N \leq L$ can be viewed as the symmetrization (over $N$ parties) of a Bell inequality $j_{N-1}\leq l$ involving only $N-1$ parties (see eq. \eqref{B_N}). It is then natural to define a frustration parameter $F = \frac{N l}{L}$. When $F=1$, there is no frustration, since there is a local deterministic model which saturates the local bound of each inequality (i.e. $j_{N-1}=l$) which thus reaches the overall upper bound of $J_N = L=Nl$. This occurs for instance in our construction of eq. \eqref{B_N}. Interestingly, this is not the case for the two inequalities above \eqref{I4}, which feature a rather large frustration: $F=11/3$ in both cases. It would be interesting to see whether large frustrations are favorable to large quantum violations.


{\bf \emph{Conclusion and Perspectives.}} We have addressed the question of distinguishing between different classes of multipartite quantum entanglement in a device-independent manner. In particular, we have presented Bell inequalities which can provably not be violated by generalized GHZ states but can be violated by W states for an arbitrary number of parties. We believe that the simplicity and robustness of these tests make them attractive from an experimental viewpoint.

Finally let us comment on some perspectives for further work. First, for $N\geq 4$, there exist many other classes of entangled states \cite{verstraete}, such as Dicke states \cite{dicke} and cluster states \cite{cluster}, for which it would be interesting to find similar tests.

Also, we believe that these ideas, in particular the concept of sub-correlation Bell inequality, may lead to further insight into multipartite quantum nonlocality. For instance, they may allow one to better understand the nonlocal correlations of the cluster state \cite{clusterNL}, which could give insight to the computational power of the latter which is still not fully understood.

\begin{acknowledgments}
	\emph{Acknowledgements:} We thank A. Acin, J.-D. Bancal, N. Gisin, Y.-C. Liang, N. Linden, M. Navascues for discussions. We acknowledge financial support from UK EPSRC and the Hungarian National Research Fund OTKA (PD101461).

\end{acknowledgments}


\section{Appendix A}
Here we derive an upper bound on the violation of the Mermin inequality $M_3\leq 2$ with the state $\ket{W_3}$. Using the method described in Ref. \cite{dimH}, based on semidefinite programming [see e.g. J. Lasserre, SIAM J. Optim. {\bf 11}, 796 (2001)], we prove that $M_3\leq 3.0792$ for $\ket{W_3}$.

We start by writing the measurement operators in the form:

\ba\label{meas} M_j = \alpha_j X + \beta_j Y + \gamma_j Z \ea
where $\alpha_j^2+\beta_j^2+\gamma_j^2=1$. Note here, since we consider a Bell inequality featuring binary outcomes and 3-qubit state, it is enough to consider qubit projective measurements.

Next, we write the density matrix $\rho_{W_3} = \ket{W_3}\bra{W_3}$ in the basis of Pauli matrices, i.e. $\{\openone, X,Y,Z\}$. We find:

\ba \rho_{W_3} &=& \frac{1}{24} (3 \openone \openone \openone + Z \openone \openone - Z Z \openone  \\\nonumber &- &  3 ZZZ  +2 (\openone +Z) (XX+YY) ) \ea
where we have omitted symmetric terms.

Our task now is compute an upper bound on the expression $E(u)=\text{tr}(\mathcal{M}_3 \rho_{W_3})$, where $\mathcal{M}_3$ denotes the Mermin Bell operator, i.e. the operator we obtain from inserting measurement operators of the form \eqref{meas} into the Mermin polynomial (see main document, eq. (5)), and $u$ denotes all the parameters characterizing the measurement operators. Thus, our problem is to solve
\ba E^* = \text{max}_u E(u)  \quad \text{s.t.} \quad c(u)\geq0 \ea
where $E(u)$ is qubic function of $u$, featuring quadratic constraints on $u$, denoted $c(u)$. This problem can be solved using semidefinite programing. Here we have used the package GloptiPoly3 [available at: http://homepages.laas.fr/henrion/software/gloptipoly3/].

We obtained $E^*\leq 3.0792$ in the second order of relaxation. This bound is most likely not tight. Numerically we have found that the optimal violation for the state $\ket{W_3}$ is $M_3\simeq 3.0460$, which leads to a resistance to noise of $w_{W_3} \simeq 0.6566$. The measurements can be chosen to be real (i.e. in the XZ plane of the Bloch sphere) and the same for all parties, that is,  of the form

\ba A_j =B_j=C_j= \cos{\theta_j} Z+ \sin{\theta_j} X .\ea
The optimal angles are given by $\theta_0=0.3002\pi$ and $\theta_1=0.8673\pi$.

As it was mentioned in the main text, the Mermin inequality is maximally violated by the GHZ state. The optimal measurements are those featured in the GHZ, namely $A_0 =B_0=C_0=X$ and $A_1 =B_1=C_1=Y$.

\section{Appendix B}

Here we give more details about the quantum violations, in particular for W states, of all the relevant inequalities presented here. In particular, we give the optimal measurement settings, which can always be taken to be real (i.e. in the XZ plane of the Bloch sphere), of the form
\ba A_j = \cos{\theta_j} Z+ \sin{\theta_j} X. \ea
Moreover, it is also optimal to consider identical measurements for all parties, i.e. $A_j=B_j=C_j$ and so on.

First, we consider the Bell inequality $B=S_3+R_2\leq 6$. The optimal violation for the state $\ket{W_3}$ is $B\simeq7.2593$, which leads to a resistance to noise of $w_{W_3} \simeq 0.8265$; the measurements are given by $\theta_0=0.2677\pi$ and $\theta_1=\pi-\theta_0$. Note that the maximal violation for qubit states is slightly larger. We have found $B\simeq 7.3084$ for the state $|\psi\rangle = 0.9971|W_3\rangle-0.07597|111\rangle$ by measurement angles $\theta_0=0.2615\pi$ and $\theta_1=\pi-\theta_0$.

Next we move to the sub-correlation Bell inequalities (see main document, eq. (15)). For inequality $I_4\leq 8$ the largest violations for the state $\ket{W_4}$ is found to be $B\simeq 11.3155$, which leads to a resistance to noise of $w_{W_4}\simeq0.7070$; the measurement settings are given by $\theta_0=0.7861\pi$ and $\theta_1=2\pi-\theta_0$. The true quantum maximum is $B\simeq 12.0680$, which occurs for the state $|\psi\rangle = 0.9877|W_4\rangle-0.1561|\overline{W}_4\rangle$ by the following measurement angles $\theta_0=0.7665\pi$ and $\theta_1=2\pi-\theta_0$, where $|\overline{W}_4\rangle$ is the spin-flipped $|W_4\rangle$ state.

For inequality $I_5\leq15$, the largest violations for the state $\ket{W_5}$ is found to be $B= 28 $, which leads to a resistance to noise of $w_{W_5}\simeq0.5357$; the measurement settings are given by $\theta_0=0$ and $\theta_1=\pi/2$. On the other hand, the true quantum maximum is $B\simeq 30.1918$, which occurs for the state $|\psi\rangle = 0.97528|W_5\rangle-0.2201|D_5^3\rangle-0.01851|11111\rangle$ by the measurement angles, $\theta_0=0$ and $\theta_1=\pi/2$, where $|D_5^3\rangle$ is the 5-qubit Dicke state with three excitations.

\section{Appendix C}

In this Appendix, we give the proof that the Bell inequality $B=S_3+R_2\leq 6$ (see main document, eq. (9)) cannot be violated by generalized GHZ states (see main document, eq. (2)) of dimension 2.

{\bf \emph{Proof.}} We start by proving the above bound for the 3-qubit GHZ, which can then be extended to all 2-dimensional generalized GHZ states at the end of the calculation.

	We first write the density matrix for the state in terms of the Pauli matrices. Using the relations
	\ba		\ket{0}\bra{0} = \frac{\openone+Z}{2}  \quad &,& \quad	\ket{1}\bra{1}  = \frac{\openone-Z}{2} \\
		\ket{0}\bra{1} = \frac{X+iY}{2} \quad &,& \quad		\ket{1}\bra{0}  = \frac{X-iY}{2} \ea	
		we have that
	\begin{align}
		\rho & = \frac{1}{2}\left(\ket{000}+\ket{111}\right) \left(\bra{000}+\bra{111}\right) \nonumber \\
		& = \frac{1}{8} \left( \openone\openone\openone + (ZZ \openone + Z\openone Z + \openone ZZ) \right. \\
		& + \left. XXX - (XYY + YXY + YYX) \right) \nonumber
	\end{align}

	Next, we wish to parameterise our measurement operators that appear in the polynomial $B$. The $m$th measurement on the qubit $j$ will be parameterised as:
	\begin{align}
		M_j^m = c_j^m XY + s_j^m Z
	\end{align}
	where $c_j^m = \mbox{cos}(t_j^m)$, $s_j^m = \mbox{sin}(t_j^m)$ and $XY$ denotes a Pauli operator on the $XY$ plane. I.e. of the form $\cos \phi X + \sin \phi Y$.

	Using this parameterisation and substituting in to $\mbox{tr}(\rho B)$ yields the value
	\begin{align}
		\mbox{tr}(\rho B) & = c_1^0 c_2^0 c_3^0 + c_1^0 c_2^0 c_3^1 + c_1^0 c_2^1 c_3^0 + c_1^1 c_2^0 c_3^0 \nonumber \\
						& - c_1^0 c_2^1 c_3^1 - c_1^1 c_2^0 c_3^1 - c_1^1 c_2^1 c_3^0 -c_1^1 c_2^1 c_3^1 \\
						& + s_1^0 s_2^1 + s_1^1 s_2^0 + s_1^0 s_3^1 + s_1^1 s_3^0 + s_2^0 s_3^1 + s_2^1 s_3^0 \nonumber
	\end{align}

	We now wish to show that this expression is upper bounded by  $6$. We will first optimise over $t_1^0$ and $t_1^1$, which may be done explicitly using the identity
	\begin{align}
		a \sin \theta + b \cos \theta = \sqrt{a^2 + b^2} \sin(\theta + \phi)
	\end{align}
	where
	\begin{align}
		\phi = \begin{cases} \arctan{(\frac{b}{a})}, & \mbox{if } a \geq 0 \\ \arctan{(\frac{b}{a})} + \pi, & \mbox{if } a < 0 \end{cases}
	\end{align}
	so the maximal value of this must be
	\begin{align}
		\mbox{max}_\theta \{ a \sin \theta + b \cos \theta \} = \sqrt{a^2 + b^2}
	\end{align}
	Hence we can write
	\begin{align}
		\mbox{tr}(\rho B) \leq \sqrt{a^2 + b^2} + \sqrt{c^2 + d^2} + s_2^0 s_3^1 + s_2^1 s_3^0
	\end{align}
	where
	\begin{align}
		a & = s_0^1 + s_3^1\\
		b & = c_2^0 c_3^0 + c_2^0 c_3^1 + c_2^1 c_3^0 - c_2^1 c_3^1 \\
		c & = s_0^0 + s_3^0\\
		d & = c_2^0 c_3^0 - c_2^0 c_3^1 - c_2^1 c_3^0 - c_2^1 c_3^1
	\end{align}
	As $s_2^0 s_3^1 + s_2^1 s_3^0$ has a maximum algebraic value of 2, it suffices to show that
	\begin{align}
		\sqrt{a^2 + b^2} + \sqrt{c^2 + d^2} \leq 4
	\end{align}
	or equivalently (by squaring the previous equation)
	\begin{align}
		a^2 + b^2 + c^2 + d^2 + 2\sqrt{(a^2 + b^2)(c^2 + d^2)} \leq 16
	\end{align}
	We will first show that
	\begin{align}
		a^2 + b^2 + c^2 + d^2 \leq 8
	\end{align}
	which can be expanded to
	\begin{align}
		a^2 + b^2 + c^2 + d^2 \leq 4 + (c_2^0 c_3^0)^2 + (c_2^1 c_3^1)^2 \nonumber \\
		+ (c_2^0 c_3^1)^2 + (c_2^1 c_3^0)^2 + 2s_2^1 s_3^1 + 2s_2^0 s_3^0 \leq 8
	\end{align}
	meaning that we only have to show
	\begin{align}
		\label{eqn-simple-2}
		(c_2^0 c_3^0)^2 + (c_2^1 c_3^1)^2 + (c_2^0 c_3^1)^2 + (c_2^1 c_3^0)^2 \nonumber \\
		+ 2s_2^1 s_3^1 + 2s_2^0 s_3^0 \leq 4
	\end{align}
	In order to do this we will find all stationary points of the function and show that at these points the function does not exceed $4$. Taking half the terms and maximising over $t_2^0$ yields
	\begin{align}
		\label{eqn-simple-1}
		\mbox{max}_{t_2^0} \{ (c_2^1 c_3^1)^2 + (c_2^0 c_3^1)^2 + 2 s_2^1 s_3^1 \} \nonumber \\
		= ((c_2^1)^2 + 1)(c_3^1)^2 + 2s_2^1 s_3^1
	\end{align}
	Differentiating this statement with respect to $t_2^1$ and setting it equal to zero allows us to find 2 conditions for a stationary point, either
	\begin{align}
		c_2^1 = 0
	\end{align}
	or
	\begin{align}
		s_3^1 = s_2^1 (c_3^1)^2
	\end{align}

	If $c_2^1 = 0$ then equation (\ref{eqn-simple-1}) simplifies to $1 \pm 2 s_3^1 - (s_3^1)^2$, which takes a maximum value when $s_3^1 = \pm 1$ and hence equation (\ref{eqn-simple-1}) $\leq 2$.
	On the other hand if $s_3^1 = s_2^1 (c_3^1)^2$ then equation (\ref{eqn-simple-1}) becomes $2(c_3^1)^2 + \tan^2(t_3^1)$, which has a stationary point only when $s_3^1 = 0$ and hence the maximum must be 2.

	The other half of equation (\ref{eqn-simple-2}) takes the same form and is proved in the same way. Therefore
	\begin{align}
		(c_2^0 c_3^0)^2 + (c_2^1 c_3^1)^2 + (c_2^0 c_3^1)^2 + (c_2^1 c_3^0)^2 \nonumber \\
		+ 2s_2^1 s_3^1 + 2s_2^0 s_3^0 \leq 4
	\end{align}

	All that remains to be done is to show that
	\begin{align}
		2\sqrt{(a^2 + b^2)(c^2 + d^2)} \leq 8
	\end{align}
	or equivalenty (again by squaring)
	\begin{align}
		(a^2 + b^2)(c^2 + d^2) \leq 16
	\end{align}

	In order to show this we will make use of the previously shown fact that $a^2+b^2+c^2+d^2 \leq 8$. Let $f = a^2+b^2$ and $g = c^2+d^2$, then we need to show that $f g \leq 16$ given that $f + g \leq 8$. This last equation can be rearranged to find
	\begin{align}
		g \leq 8 - f \Rightarrow fg \leq f(8-f)
	\end{align}
	which is maximised when $f=4$, which means that the maximum value of $fg$ must be $16$. This final part proves that $\mbox{tr}(\rho B) \leq 6$ for GHZ states.

	We now extend this result to all generalized GHZ states of dimension 2:
	\begin{align}
		\rho & = \left( c_\theta \ket{000} + s_\theta \ket{111} \right) \left( c_\theta \bra{000} + s_\theta \bra{111} \right) \nonumber \\
			& = \frac{1}{8} \left[ \openone \openone \openone  + (ZZ \openone +Z\openone Z+\openone ZZ) \right. \\
			& + c_{2\theta}(\openone Z\openone +Z\openone \openone +\openone \openone Z+ZZZ) \nonumber \\
			& \left. + s_{2\theta}(XXX-(XYY+YXY+YYX)) \right] \nonumber
	\end{align}
	Therefore we can calculate the value of $\mbox{tr}(\rho B)$ using our previous measurement parameterisation
	\begin{align}
		\mbox{tr}(\rho B) & = s_{2\theta}\left( c_1^0 c_2^0 c_3^0 + c_1^0 c_2^0 c_3^1 + c_1^0 c_2^1 c_3^0 + c_1^1 c_2^0 c_3^0 \right. \nonumber \\
						& - \left. c_1^0 c_2^1 c_3^1 - c_1^1 c_2^0 c_3^1 - c_1^1 c_2^1 c_3^0 -c_1^1 c_2^1 c_3^1\right) \\
						& + s_1^0 s_2^1 + s_1^1 s_2^0 + s_1^0 s_3^1 + s_1^1 s_3^0 + s_2^0 s_3^1 + s_2^1 s_3^0 \nonumber
	\end{align}
	It is possible to maximise this statement over $\theta$ analytically by finding the function's stationary points. Differentiating and setting it equal to zero means either
	\begin{align}
		\label{eqn-cosines}
		c_1^0 c_2^0 c_3^0 + c_1^0 c_2^0 c_3^1 + c_1^0 c_2^1 c_3^0 + c_1^1 c_2^0 c_3^0 \nonumber \\
		- c_1^0 c_2^1 c_3^1 - c_1^1 c_2^0 c_3^1 - c_1^1 c_2^1 c_3^0 -c_1^1 c_2^1 c_3^1 = 0
	\end{align}
	or
	\begin{align}
		\label{eqn-cos-theta}
		\cos(2\theta) = 0
	\end{align}
	If equation (\ref{eqn-cosines}) is true then the equation reduces to a sum of 6 values, each less than or equal to 1, so the bound holds. If equation (\ref{eqn-cos-theta}) holds then $\theta=\frac{\pi}{4}$ and the state reduces to a GHZ state, where we have already proven the bound. Therefore this bound of 6 holds for all Schmidt states as well.

\end{document}